\documentclass{jfm}
\usepackage{graphicx}
\usepackage{epstopdf, epsfig}
\usepackage{url} 
\usepackage[colorlinks, allcolors=blue]{hyperref}

\usepackage{natbib}
\usepackage[colorlinks]{hyperref}
\usepackage{etoolbox}
\usepackage{soul}

\makeatletter

\patchcmd{\NAT@citex}
  {\@citea\NAT@hyper@{%
     \NAT@nmfmt{\NAT@nm}%
     \hyper@natlinkbreak{\NAT@aysep\NAT@spacechar}{\@citeb\@extra@b@citeb}%
     \NAT@date}}
  {\@citea\NAT@nmfmt{\NAT@nm}%
   \NAT@aysep\NAT@spacechar\NAT@hyper@{\NAT@date}}{}{}

\patchcmd{\NAT@citex}
  {\@citea\NAT@hyper@{%
     \NAT@nmfmt{\NAT@nm}%
     \hyper@natlinkbreak{\NAT@spacechar\NAT@@open\if*#1*\else#1\NAT@spacechar\fi}%
       {\@citeb\@extra@b@citeb}%
     \NAT@date}}
  {\@citea\NAT@nmfmt{\NAT@nm}%
   \NAT@spacechar\NAT@@open\if*#1*\else#1\NAT@spacechar\fi\NAT@hyper@{\NAT@date}}
  {}{}

\makeatother

\shorttitle{Magnetohydrodynamic Hele-Shaw Flows}
\shortauthor{K. I. McKee}

\title{Exact and Approximate Solutions for Magnetohydrodynamic Flow Control in Hele-Shaw Cells}

\author{Kyle I. McKee\aff{1,2}
  \corresp{\email{kimckee@mit.edu}}
  }

\affiliation{\aff{1}Department of Mathematics, Massachusetts Institute of Technology,
Cambridge, MA 02139, USA
\aff{2}Gulliver Laboratory,
ESPCI Paris,
PSL University,
75231 Paris cedex 05, France
}

\begin{document}

\maketitle
\begin{abstract}
Consider the motion of a thin layer of electrically conducting fluid, between two closely spaced parallel plates, in a classical Hele-Shaw geometry. Furthermore, let the system be immersed in a uniform external magnetic field (normal to the plates) and let electrical current be driven between conducting probes immersed in the fluid layer. In the present paper, we analyse the ensuing fluid flow at low Hartmann numbers.
We first elucidate the mechanism of flow generation both physically and mathematically. We proceed by presenting mathematical solutions for a class of canonical multiply-connected geometries, in terms of the prime function developed by \cite{crowdy2020solving}. Notably, those solutions can be written explicitly as series, and are thus exact, in doubly-connected geometries. Note that in higher connectivities, the prime function must be evaluated numerically.
We then demonstrate how recently developed fast numerical methods may be applied to accurately determine the flow-field in arbitrary geometries when exact solutions are inaccessible.

\end{abstract}

\section{Introduction}\label{intro}
Magnetohydrodynamic flows find applications in a variety of fields from geophysics and astrophysics to metallurgy, in situations where the fluid under study interacts significantly with external or self-induced electromagnetic fields \citep{moffatt1978field,davidson2002introduction}.
Although magnetic effects tend to transform the Navier-Stokes equations into an even more formidable form, they also generate a host of unique phenomena that have been studied over the past century including Alfv{\'e}n waves \citep{alfven1942existence} and geodynamos \citep{moffatt1978field,moffatt2019self}. In the present paper, we analyse mathematically the manner in which electrical and magnetic effects modify arguably the simplest possible base flow — a two-dimensional potential flow as is physically realized in a Hele-Shaw cell. 

First, consider a Hele-Shaw flow in the absence of electromagnetic effects. The flow is obtained through the solution of a two-dimensional boundary-value problem for a harmonic pressure-field whose negative gradient gives the velocity field. Since the pressure field must be single-valued, the fluid flow must possess zero circulation, thus removing the usual paradox of undetermined circulations present in aerofoil theory (see \citet{gonzalez2022variational} for a recent development and survey of the aerofoil problem). Since pressure-driven Hele-Shaw flows possess exactly zero circulation around any closed contour, circulation cannot be used to mix or control such flows. Since the Hele-Shaw cell constitutes a model for flows through porous media and microfluidic devices, it is of technological interest to overcome the no-circulation limitation. We now outline some relevant literature regarding circulation generation via electromagnetic effects.

\cite{moffatt1991electromagnetic} discussed mechanisms for stirring using time-dependent electromagnetic fields, in general three-dimensional contexts. The mechanisms presented therein rely mainly on truly three-dimensional aspects of flow, and the Hele-Shaw limit was not considered. 
Recently, \citet{mirzadeh2020vortices} showed that circulation can be generated in electro-osmotic  Hele-Shaw flows if the gap thickness is made inhomogeneous. Henceforth, when discussing Hele-Shaw cells, we restrict our attention to the case of uniform thickness.

\cite{bau2001minute} and \cite{zhong2002magneto} fabricated Hele-Shaw type devices containing thin layers of electrolytes, through which electrical current was driven between electrodes immersed in the fluid. The former work placed electrodes on the base of the device whereas the latter placed electrodes on the walls of a concentric annlulus. When the devices were placed in a uniform magnetic field, Lorentz forces induced a fluid flow in the bulk. \cite{yi2002magnetohydrodynamic} showed that under certain conditions, the application of periodic AC electrical currents in similar devices may lead to chaotic advection.
Later, \cite{homsy2005high} manufactured a high current density DC microfluidic pump which found applications in nuclear magnetic resonance \citep{homsy2007magnetohydrodynamic}; notably, the pump avoided significant Joule heating. A recent comprehensive review of applications of magnetohydrodynamics in the context of microfluidics is given by \cite{bau2022applications}. 

The papers discussed in the paragraph above were mainly experimental in nature. In the cases where fluid flows were analysed analytically, such as in \cite{bau2001minute} and \cite{zhong2002magneto}, solutions were obtained as infinite series in highly specialized coordinate systems, using methods only applicable to special geometries. One aim of the present paper is to demonstrate how conformal maps obviate these complicated analyses and yield a class of flow geometries with closed-form solutions; in doubly-connected problems, the prime function has a series representation and solutions are exact \citep{baddoo2020exact}.

More recently, \cite{david2023magnetostokes} established a mathematical analogy, in magneto-hydrodynamics, to reversible Stokes flow in a Taylor-Couette geometry.
Experiments presented therein beautifully reproduced the classic kinematic reversibility experiments of \cite{kinrev} without the need for moving boundaries. At the end of their paper, the authors noted that in the limit of a shallow fluid layer, the flow resembles potential flow. Note that their experiments were conducted with a free surface and no lid. Although the authors correctly pointed out the potential flow limit, no analytical solutions or further theoretical developments were explored in their work.

\begin{figure}
  \centerline{\includegraphics[width=0.95 \textwidth]{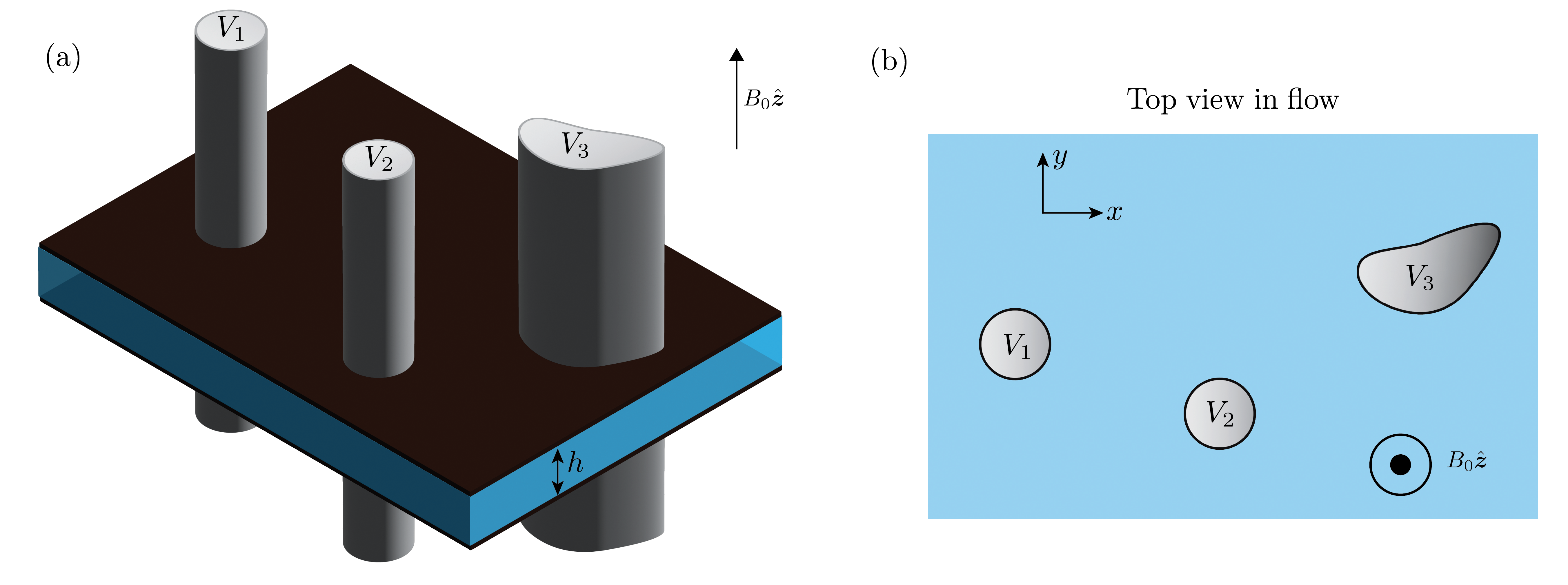}}
  \caption{(a) Schematic of the system under consideration in the present paper. Voltages are applied to conducting bodies immersed in a thin layer of conducting fluid, of thickness $h$, that is bounded above and below by parallel walls. The entire system is immersed in a uniform magnetic field oriented along the $z$-axis and normal to the bounding walls. (b) Top-view of the flow geometry. Each conducting probe serves as an impermeable obstacle in the fluid flow. Each probe is held at a fixed voltage. We also investigate the effect of insulating obstacles in the flow, which obey the zero Neumann condition, $\nabla V \cdot \boldsymbol{n}=0$, in place of the Dirichlet condition satisfied by conductors, where $\boldsymbol{n}$ is the unit normal vector to the surface of the obstacle.}
\label{fig:schem}
\end{figure}

In the present paper, we analyse generally the flow of a conducting fluid in a Hele-Shaw cell, immersed in an external uniform and constant magnetic field that is directed normal to the cell walls, $\boldsymbol{B}=B_0\hat{\boldsymbol{z}}$ (see figure \ref{fig:schem}). We consider scenarios where flow is driven by applying voltages to conducting probes that are immersed in the fluid, the probes being impermiable to fluid flow. We assume an ohmic fluid to model the electrical current flow, $\boldsymbol{J}=\sigma \boldsymbol{E}$, where $\sigma$ is the electrical conductivity of the fluid and $\boldsymbol{E}$ is the electric field experienced by the fluid. The presence of electrical current leads to Lorentz forces acting on the fluid bulk, $\boldsymbol{F}=\boldsymbol{J}\times\boldsymbol{B}$, which ultimately induces a fluid flow. We also consider the effect of adding obstacles to the flow that are either electrical insulators or conductors. We present mathematical solutions for a large class of multiply-connected geometries in terms of the prime function given by \cite{crowdy2020solving}. In general geometries where exact solutions are not possible, we show how highly accurate approximate solutions can be obtained using series solution methods as described by \cite{trefethen2018series}.



 The remainder of this paper is arranged as follows.  In \S \ref{physpic}, we discuss the simplifying assumptions in our mathematical formulation, and subsequently outline the physical mechanism driving the fluid flow.  In \S \ref{solution}, we present a complex variables formulation of the model described in \S \ref{physpic}. We proceed by deriving mathematical solutions for the fluid flow in a variety of multiply-connected geometries, using the framework of \cite{crowdy2020solving}. One such solution gives the flow in a geometry explored experimentally by \cite{david2023magnetostokes}. 
 In \S \ref{exp}, another solution is compared to a new experiment which serves as further motivation for the present theoretical study.  In \S \ref{num}, we show how more general geometries can be solved to high accuracy using series solutions.
 
\section{Assumptions and Physical Picture}\label{physpic}
Consider a thin layer of conducting fluid occupying the region between two rigid walls separated by a distance $h$, as in figure \ref{fig:schem}(a). The conducting fluid may be taken, for example, to be saltwater or liquid mercury. Perfectly conducting probes, held at fixed voltages, penetrate the entire cell thickness $h$. Meanwhile, the system is immersed in an external magnetic field perpendicular to the flow, $\boldsymbol{B}=B_0\hat{\boldsymbol{z}
}$. The application of a voltage difference between two conductors gives rise to an electric field in the conducting fluid, which induces an electrical current flow according to Ohm's law. The external magnetic field then induces a Lorentz force on fluid parcels which gives rise to a fluid flow whose direction can be predicted using the right-hand rule.

Restricting our attention to steady and quasi-steady problems, the voltage $V(\boldsymbol{x})$ is determined generally as the solution to a Poisson equation. In an uncharged ohmic material, the Poisson equation reduces to the Laplace equation so that $V(\boldsymbol{x})$ must be a harmonic function. We now justify the proposed physical picture. 

In the studies mentioned in \S \ref{intro}, as well as our experiment in \S \ref{exp}, the conducting fluid is a binary electolyte and so the ``uncharged" assumption requires extra justification. A binary electrolyte comprises two oppositely charged species, whose concentrations we denote $c^+$ and $c^-$. If at any point of space $c^+\neq c^-$ holds, then the Poisson equation does not reduce to the Laplace equation, since the charge density (source term) is not identically zero in all space. In such a case, $V(\boldsymbol{x})$ fails to be harmonic across the entire domain. To ensure that $V(\boldsymbol{x})$ is indeed harmonic, we must invoke the so-called local electroneutrality condition, $c^+ = c^-$, which is a standard assumption in the modelling of electrolytes \citep{zaltzman2007electro}. This assumption is valid in the bulk of the fluid but is violated in the electric double layer, within a Debye length of conducting boundaries. In the present study, we make no effort to model double layer effects. Instead we treat boundaries as simplified ideal conductors and insulators, and the bulk fluid as a locally electroneutral ohmic conductor. We now proceed under the assumption that the electrical potential can be modelled as a harmonic function in the fluid domain.

Under the stated assumptions, the electric field as measured in the lab frame is given by the gradient of a harmonic potential, $\boldsymbol{E}_{\mathrm{lab}}=-\nabla V$. The electric field drives an electrical current according to Ohm's law, $\boldsymbol{J}=\sigma \boldsymbol{E}$, where $\boldsymbol{E}$ is the electric field felt in the frame of a fluid parcel. 
The electric field that a fluid parcel experiences is related to the electric field in the lab frame by the relation $\boldsymbol{E}=\boldsymbol{E}_{\mathrm{lab}}+\boldsymbol{u}\times \boldsymbol{B}$, owing to the fact that the electric field is not invariant under Gallilean transformations (see \citet[pp. 34]{moffatt1978field}). However, we will now show that the second term is negligible in the context of the Hele-Shaw flow under consideration here. In our system, typical voltage differences between probes are $1V$; voltages higher than roughly $1.23 V$ create bubbles due to water electrolysis. With typical separation distances between electrical probes being on the order of centimeters, the typical electric field magnitude is $E_0\approx1V/\mathrm{cm}$. Typical velocities in the cell are $U_0\approx1\mathrm{mm}/\mathrm{s}$, while the maximum magnetic field is roughly $B_0 \approx 2340 \mathrm{G}$. Hence, the relative importance of $\boldsymbol{u}\times \boldsymbol{B}$ as compared to $\boldsymbol{E}_{\mathrm{lab}}$ scales as $U_0B_0/E_0=\mathcal{O}(10^{-6})$ and we thus safely neglect the cross product term. Henceforth, we use $\boldsymbol{E}$ to denote the electric field and ignore any distinction between the field experienced in different frames.

The equations of conservation of fluid momentum and mass then become,
\begin{subeqnarray}\label{eq:ns1}
    \rho\left(\frac{\partial \boldsymbol{u}}{\partial t}+\boldsymbol{u}\boldsymbol{\cdot}\nabla \boldsymbol{u}\right) & = &\mu \nabla^2 \boldsymbol{u}-\nabla P+\sigma \boldsymbol{E}\times \boldsymbol{B}, \\[3pt]
    \nabla \boldsymbol{\cdot} \boldsymbol{u} & = &  0.
\end{subeqnarray}

Since the cell thickness, $h$, is much smaller than the lateral extent of the system, the Hele-Shaw approximation is justified, implying that viscous forces dominate inertial forces (see \citet[pp. 222]{batchelor1967introduction}). Hence, the left side of (\ref{eq:ns1}a) can be neglected, and the flow is determined by a balance of viscous forces and both pressure and magnetic driving forces. In order to invoke the usual Hele-Shaw approximation, it is important to ensure that the flow across the entire gap thickness, $h$, is fully developed. The presence of a magnetic field has the ability to shrink the boundary layer so that the fully-developed parabolic velocity profile may not be reached \citep[ figure 5.10 on pp. 153]{davidson2002introduction}; see also the work of \citet{rossow1958flow} for an interesting application of this concept. In our system, the Hartmann number is small, $Ha=\sqrt{B_0^2 h^2 \sigma/\mu }\approx 0.01$, indicating that magnetic dissipation is neglibible and the flow does indeed adopt the fully-developed viscous parabolic profile; our calculations use the typical gap thickness, $h\approx 0.7 \mathrm{mm}$ (see \S \ref{exp}). For comparison, liquid mercury in a Hele-Shaw cell of the same thickness, and subject to the same magnetic field, has a larger Hartman number $Ha\approx 4$; thus, a thinner gap is necessary to attain the Hele-Shaw limit ($Ha \ll 1$) for the liquid metal.

Under the Hele-Shaw approximation, the top-down velocity becomes two-dimensional and we can immediately write
\begin{subeqnarray}\label{eq:hs1}
    \boldsymbol{u} & = &\frac{1}{2\mu}\left( -\nabla P+\sigma B_0\boldsymbol{E}\times \hat{\boldsymbol{z}}\right)z(h-z), \\[3pt]
    \nabla \boldsymbol{\cdot} \boldsymbol{u} & = &  0.
\end{subeqnarray}

To make further progress, we note that the Lorentz force, 
$\boldsymbol{F}=-\sigma B_0 \nabla V \times  \hat{\boldsymbol{z}}$, is clearly solenoidal and irrotational so that $\boldsymbol{\nabla}\boldsymbol{\cdot}\boldsymbol{F}=0$ and $\boldsymbol{\nabla}\times\boldsymbol{F}=0$. It is therefore possible to represent it as the gradient of a harmonic function, $\boldsymbol{F}=\nabla \phi$, where $\nabla^2\phi=0$. In general $\phi$ need not be single-valued.
Equation (\ref{eq:hs1}a) then reduces to
\begin{equation}\label{eq:floweq}
    \boldsymbol{u}=\frac{z(h-z)}{2\mu}\nabla\left(\phi(x,y)-P(x,y)\right),
\end{equation}
where the gradient is two-dimensional. Henceforth, we shall suppress the vertical structure of the flow, $z(h-z)$, for convenience since we are only concerned with a top-down view of the flow. 
Note that since $\phi$ is a harmonic function, (\ref{eq:hs1}b) and (\ref{eq:floweq}) together imply that $P$ must be harmonic too.

While $P$ must be a single-valued function, there is no such restriction on $\phi$. In fact, the main manner in which the Lorentz force generates flow, as will be shown, is by inducing flow circulation which actually requires $\phi$ to be multi-valued. Physically, the circulation is induced as follows. Since electric field lines must exit perpendicular to conducting surfaces, so too must the current density, $\boldsymbol{J}$, according to Ohm's law. The right-hand rule then reveals that the Lorentz force induces a circulatory flow around each conductor.




\section{Complex Variables Formulation and Solution Procedure}\label{solution}
We proceed by formulating the boundary value problem for the two-dimensional velocity field, $\tilde{\boldsymbol{u}} \equiv \nabla (\phi - P)$, using a complex variables approach. The solution procedure for obtaining $\tilde{\boldsymbol{u}}$ is as follows. First, one must solve the electrostatic problem for $V(x,y)$ in the fluid domain. Once $V$ is known, the potential $\phi$, which describes the Lorentz force according to $\boldsymbol{F}=\nabla \phi$, may be obtained. Finally, a single-valued pressure field $P$, as appears in (\ref{eq:floweq}), must be obtained to enforce the impermeability condition on the surface of each flow obstacle. The procedure is outlined in detail in the remainder of \S \ref{solution}.

\subsection{Complex Variables Formulation}\label{compvar}
Consider the electrostatic boundary value problem in the domain illustrated in figure \ref{fig:schem}(b). We seek a real harmonic function $V(x,y)$ in the fluid domain satisfying constant Dirichlet conditions on each conducting boundary. More generally, we consider also the addition of perfectly insulating obstacles, on the surface of which $V$ satisfies a zero Neumann boundary condition.

On each conducting surface, $\partial B_i$, constant Dirichlet boundary data is prescribed so that $V=V_j$ for $j\in\{1,2,..., N_C\}$, where $N_C$ is the total number of conductors. On insulating boundaries, $\partial D_k$, the normal derivative is required to vanish so that $\nabla V \cdot \boldsymbol{n}=0$ where $k\in\{1,2,..., N_I\}$ and $N_I$ is the total number of electrical insulators. To make progress, we now express the boundary value problem in the language of complex variables.

We may take $V$ to be the real part of an analytic function $W_E(z)$, where $z=x+iy$. We require $W_E(z)$ to be analytic over the entire fluid domain to ensure the harmonicity of $V(x,y)$. We thus seek an analytic function $W_E$ with the properties
\begin{subeqnarray}\label{eq:Webvp}
    \mathrm{Re}\left\{W_E(z)\right\} & = &V_j\;,\; z \in \partial B_j\;, j\in\{1,2,..., N_C\},\\[3pt]
    \mathrm{Im}\left\{ W_E(z)\right\} & = & C_j\;,\; z \in \partial D_j\;, j\in\{1,2,..., N_I\},
\end{subeqnarray}
where $C_j$ are real constants that are unknown a priori. In (\ref{eq:Webvp}b), the zero Neumann condition on $V$ has been re-expressed as a constant condition on its harmonic conjugate.

Suppose now that $W_E$, satisfying (\ref{eq:Webvp}), is known. Then the complex electric field is given by $E\equiv E_x+\mathrm{i}E_y=-\overline{dW_E/dz}$, from which the current may be obtained via Ohm's law, $J=-\sigma\overline{dW_E/dz}$. The cross product giving the Lorentz force is performed through a pre-multiplication by $-\mathrm{i}$, which corresponds to a rotation through an angle of $-\pi/2$ in the plane, giving $F=\mathrm{i}\sigma B_0\overline{dW_E/dz}$. The Lorentz potential and force are then written as
\begin{subeqnarray}\label{eq:FDEF}
    W_L & = & -\mathrm{i}\sigma B_0 W_E \\[3pt]
    F & = & \overline{dW_L/dz}.
\end{subeqnarray}
In complex notation, the two-dimensional velocity $\tilde{\boldsymbol{u}}$ may thus be written as
\begin{equation}\label{eq:floweq_comp}
    u= \overline{ \frac{d}{dz}\left(W_L-W_P\right) }\equiv \overline{\frac{d}{dz}\left(W_{\mathrm{flow}}\right) },
\end{equation}
where $W_P$ is an analytic function satisfying $\mathrm{Re}\left\{W_P\right\}=P(x,y)$. In (\ref{eq:floweq_comp}), we have defined the flow potential function $W_{\mathrm{flow}}=W_L-W_P$.

The precise mechanism for the generation of circulation is now evident and can be described as follows. First, the electrostatic problem (\ref{eq:Webvp}) is solved in the fluid domain exterior to the collection of insulators and conductors held at different electrical potentials. Depending on the geometry and the applied voltages, some amount of surface charge, $Q_j$, accumulates on each conductor surface $\partial B_j$ in the electrostatic problem. Each such surface charge manifests as a source term in the complex potential $W_E$ so that the potential satisfies the relation,
\begin{equation}\label{eq:WEsourcebehav}
W_E = -\sum_{j=1}^{N_C}\frac{Q_j}{2\pi} \log{\left(z-z_j\right)}+\mathrm{single\;valued\;function},
\end{equation}
where $z_j \in B_j$.
The amount of current, $I_j$, leaving each conductor, $B_j$, is then trivially related to the induced charge in the electrostatic problem through Ohm's law so that $I_j=\sigma Q_j$.

The Lorentz force potential is obtained through (\ref{eq:FDEF}a), which converts the source term present in (\ref{eq:WEsourcebehav}) into a circulation term through a pre-multiplication by $\mathrm{i}$. Since the pressure is a single-valued function, we immediately deduce that
\begin{equation}\label{eq:Wfform}
W_{\mathrm{flow}}=-\mathrm{i}\frac{ B_0}{2\pi}\sum_{j=1}^{N_C} I_j  \log{\left(z-z_j\right)}+W_{\mathrm{flow}}^{\mathrm{SV}},
\end{equation}
where $W_{\mathrm{flow}}^{\mathrm{SV}}$ is a single-valued function. Thus, the amount of electrical current leaving a particular conductor alone dictates the induced circulation around that same conductor in the induced fluid flow, $\Gamma_j=B_0 I_j$.

After obtaining $Q_i$ through the solution to the electrostatic problem (\ref{eq:Webvp}), the boundary value problem for $W_{\mathrm{flow}}$ becomes fully specified by (\ref{eq:Wfform}) supplemented by impermeability conditions 
on each conducting body which may be expressed as follows,
\begin{equation}\label{eq:Wfbvp}
    \mathrm{Im}\left\{W_{\mathrm{flow}}^{\mathrm{SV}}(z)\right\}  = \mathrm{Im}\left\{\mathrm{i}\frac{ B_0}{2\pi}\sum_{j=1}^{N_C} I_j  \log{\left(z-z_j\right)}\right\}+\psi_k,
\end{equation}
where $z \in \partial B_k$ for $k\in\{1,2,..., N_C\}$, and $\psi_k$ are real constants that are unknown apriori. We have assumed the absence of additional driving forces. However, we note that a background free stream of complex velocity $U$, for example, may be included simply by augmenting the right side of (\ref{eq:Wfform}) with the term $\overline{U}z$ and the right side of (\ref{eq:Wfbvp}) with the term $-\mathrm{Im}\left\{\overline{U}z\right\}$.

Note that even in the case of a strictly magnetically driven flow, a non-zero pressure field ($W_P \neq 0$) may be required to enforce the impermeability boundary conditions, to be explained as follows. In the case that all obstacles in the flow are perfect conductors ($N_I=0$), it is clear that $W_{\mathrm{flow}}=W_L=-\mathrm{i}\sigma B_0 W_E$ alone satisfies the fluid flow boundary conditions and $W_P=0$. Since the electric field lines of $W_E$ are necessarily perpendicular to the boundary of a perfect conductor, the multiplication by $\mathrm{i}$ in (\ref{eq:FDEF}a) ensures that the fluid flow streamlines are tangent to each conducting surface and thus $W_L$ is the valid fluid flow potential .

However, in cases where some obstacles are electrical insulators, $W_L$ alone does not satisfy the impermeability boundary conditions. On the surface of each insulator, the electric field lines of $W_E$ are necessarily tangent to the insulating surface. Hence, the multiplication by $\mathrm{i}$ in (\ref{eq:FDEF}a) produces fluid flow streamlines that are normal to each insulating surface, so that $W_L$ alone violates the impermeability condition on insulators. Hence, when insulators exist in the flow, a non-zero pressure field develops to enforce the impermeability condition on insulators. When electrically insulating obstacles are present, the fluid flow must thus be obtained in two steps. First, one must solve the electrostatic problem and thus obtain a set of charges $Q_j$ accumulated on the surface of each conducting body. Second, those charges are used to specify the circulation in the fluid flow problem so that the flow problem described by (\ref{eq:Wfform})-(\ref{eq:Wfbvp}) is fully posed and can be solved. We note again that since the circulation around each body is specified, the problem is well-posed and we do not encounter the issue of undetermined circulations that plagues high Reynolds number aerofoil theory \citep{gonzalez2022variational}. We proceed in \S \ref{conformalin} with a brief discussion relating the electrostatic problem (\ref{eq:Webvp}) to the induced circulations. We subsequently derive mathematical solutions for the flow in various multiply-connected geometries.

\subsection{Connection Between Induced Circulation Electrostatic Capacitance}\label{conformalin}
It is worth noting that in the special case where only two distinct voltage values are prescribed on the boundary, (\ref{eq:Webvp}) is an electrostatic capacitance problem. The electrostatic capacitance is the measure of the magnitude of induced charge on each conducting surface per unit applied voltage difference, and it is determined by the geometry. It is a conformal invariant and is the reciprocal of the extremal distance \citep[pp. 65]{ahlfors2010conformal}. A significant literature exists regarding bounds and comparison theorems for the capacitance of different geometries \citep[pp. 65]{ahlfors2010conformal}. Recall that in \S \ref{compvar}, it was shown that the circulation of the induced fluid flow is solely determined by the charges present in the electrostatic problem (\ref{eq:Webvp}). Thus, theorems which indicate how one might geometrically tune the capacitance of the electrostatic problem also indicate how one might tune the circulation of the induced fluid flow.
For example, the introduction of an insulator anywhere into a given probe configuration necessarily decreases the capacitance and hence also the induced circulation. Meanwhile, the introduction of a floating conductor necessarily increases the capacitance and circulation.



\subsection{Magnetically-Driven Flows Around a Collection of Conductors ( $W_P=0$ )}\label{voltonly}
\begin{figure}
  \centerline{\includegraphics[width=0.95 \textwidth]{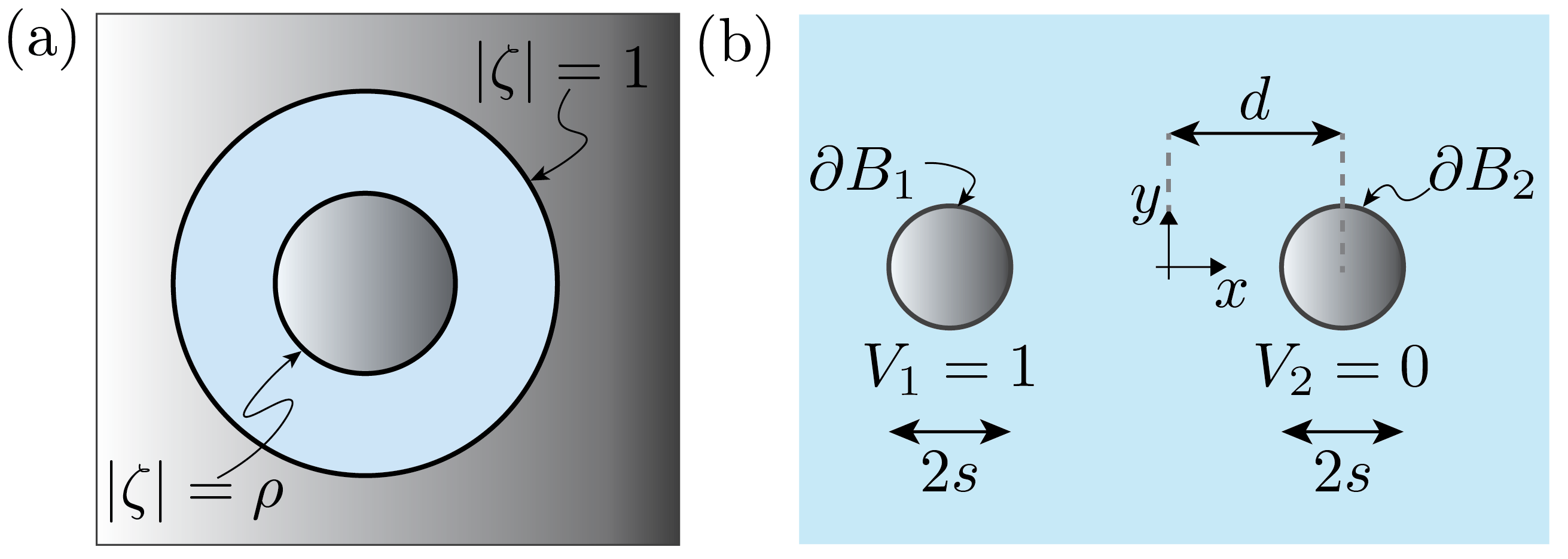}}
  \caption{(a) Conformally mapped domain of the physical geometry given in panel (b). The circles $|\zeta|=1$ and $|\zeta|=\rho$ map to the two conductor boundaries in the physical plane. (b) Schematic for electrostatic problem between two perfectly conducting cylinders held at fixed voltages in the physical $z$-domain.}
\label{fig:confmap}
\end{figure}

Through example, we now outline a procedure for obtaining solutions when there are no electrically insulating obstacles in the flow ($N_I=0$). We consider the geometry in figure \ref{fig:confmap}(b), which is a special case of the general geometry of figure \ref{fig:schem}(b) where $N_I=0$ and $N_C=2$.
Two conducting obstacles lie in a two-dimensional conducting fluid. The first is held at the voltage $V_1=1$ while the second is held at $V_2=0$. Otherwise the flow is infinite in extent and there are no external pressures driving fluid flow. In order to determine the induced fluid flow, we first must solve (\ref{eq:Webvp}) for the electrical potential, and then the resulting flow is given by (\ref{eq:FDEF}a) and (\ref{eq:floweq_comp}) after taking $W_P=0$, as was outlined in \S \ref{compvar}, since all the immersed bodies are perfect conductors. We solve for $W_E$ by first conformally mapping the physical domain in figure \ref{fig:confmap}(b) onto the concentric annulus of figure \ref{fig:confmap}(a).

The conformal map from the physical geometry in figure \ref{fig:confmap}(b) to the an annulus of figure \ref{fig:confmap}(a) is given by, 
\begin{equation}\label{eq:map}
\zeta(z)=\sqrt{\rho}\frac{\sqrt{d^2-s^2}-z}{\sqrt{d^2-s^2}+z},
\end{equation}
where the inner annular radius is given by $\rho=\left( \left( d-\sqrt{d^2-s^2} \right)/s \right)^2$. Note that for unit circles, $s=1$. The left conductor in the physical domain is mapped to the inner circle $|\zeta|=\rho$ while the right conductor is mapped the the outer circle of the annulus, $|\zeta|=1$. It is simple to solve the Dirichlet problem in the conformally mapped domain. In order to achieve $V=1$ on $|\zeta|=\rho$ and $V=0$ on $|\zeta|=1$, the potential in the annulus must be
\begin{equation}\label{eq:elecpot}
W_{E,\zeta}(\zeta)=V_0\frac{\log{\left(\zeta/\rho\right)}}{\log{\left(1/\rho\right)}}.
\end{equation}
The complex potential in the physical domain is then given simply by 
\begin{equation}
    W_E(z)=W_{E,\zeta}(\zeta(z))=V_0 \frac{\log{\left(\frac{1}{\sqrt{\rho}}\frac{\sqrt{d^2-s^2}-z}{\sqrt{d^2-s^2}+z}\right)}}{\log{\left(1/\rho\right)}}.
\end{equation}
The complex potential for the flow is then obtained simply through multiplication by $-\mathrm{i} \sigma B_0$, as follows
\begin{equation}\label{eq:lorpot}
    W_{\mathrm{flow}}(z)=-\mathrm{i} \sigma B_0V_0 \frac{\log{\left(\frac{1}{\sqrt{\rho}}\frac{\sqrt{d^2-s^2}-z}{\sqrt{d^2-s^2}+z}\right)}}{\log{\left(1/\rho\right)}},
\end{equation}
It is clear from (\ref{eq:lorpot}) that the applied voltage difference manifests as a fluid flow circulation of magnitude $\Gamma = 2\pi\sigma B_0 V_0$ around each of the cylinder conductors. The induced fluid flow is plotted in figure \ref{fig:condcase}(a).
\subsection{Uniform Stream Past Perfect Conductors}
We now derive the solution for a strictly pressure-driven flow — which possesses zero circulation — past the same two cylinders. We proceed by demonstrating, through direct calculations, how magnetic fields may induce circulation in the pressure-driven flow.
\subsubsection{The Pressure-Driven Uniform Stream (No Circulation)}\label{unifnocurr}
Consider again the circular obstacle geometry given in figure \ref{fig:confmap}(b), except now a uniform steam $U\in \mathbb{R}$ is directed along the real axis past the two cylinders with zero circulation around each cylinder. Such a flow might be generated, for example, by an applied pressure gradient along the $\hat{\boldsymbol{x}}$ direction. 

In either case, the flow solution can be obtained directly by the methods of \citet[Chapter 15]{crowdy2020solving}, who presents a calculus for potential theory. With vanishing circulations around each body, and the behaviour at infinity specified ($W_{_{\mathrm{flow}}}\sim U z$ as $|z|\rightarrow \infty$), a class of mathematical solutions for the magnetohydrodynamically driven flow in the Hele-Shaw cell become expressible in terms of the prime function. Notably, the prime function has a closed-form series reprentation in the doubly-connected case \cite{baddoo2020exact}. In what follows, we present the solution procedure for the two-cylinder problem. We later outline how N-body solutions follow by an identical procedure (for more details, see \citet[pp. 294]{crowdy2020solving}).

Consider two unit cylinders a distance $2d$ apart, with each centered on the real axis, as in figure \ref{fig:confmap}(b). Take the left-most circle to be centered on the origin. The M{\"o}bius transformation $\zeta=1/z$ then maps the origin-centered circle to itself. However, the second circle is mapped to a circle of radius $\delta<1$ inside the unit circle and centered at the point $q$. More generally, if additional cylinders were added to the physical domain, the same M{\"o}bius transformation would take each additional cylinder to a distinct excised circle contained inside the unit disk. The unit circle, with a number of excised interior circles, represents a canonical domain, where the so-called prime function $\omega(z_1,z_2)$ becomes useful for obtaining mathematical solutions to certain boundary value problems \citep{crowdy2020solving}. Let us return to the two-cylinder problem of current interest. In this case (figure \ref{fig:confmap}(b)), it is convenient to choose the canonical domain to be a concentric annulus ($q=0$) as drawn in figure \ref{fig:confmap}(a), since the solution possesses an explicit sum representation there.

\begin{figure}
  \centerline{\includegraphics[width=0.95 \textwidth]{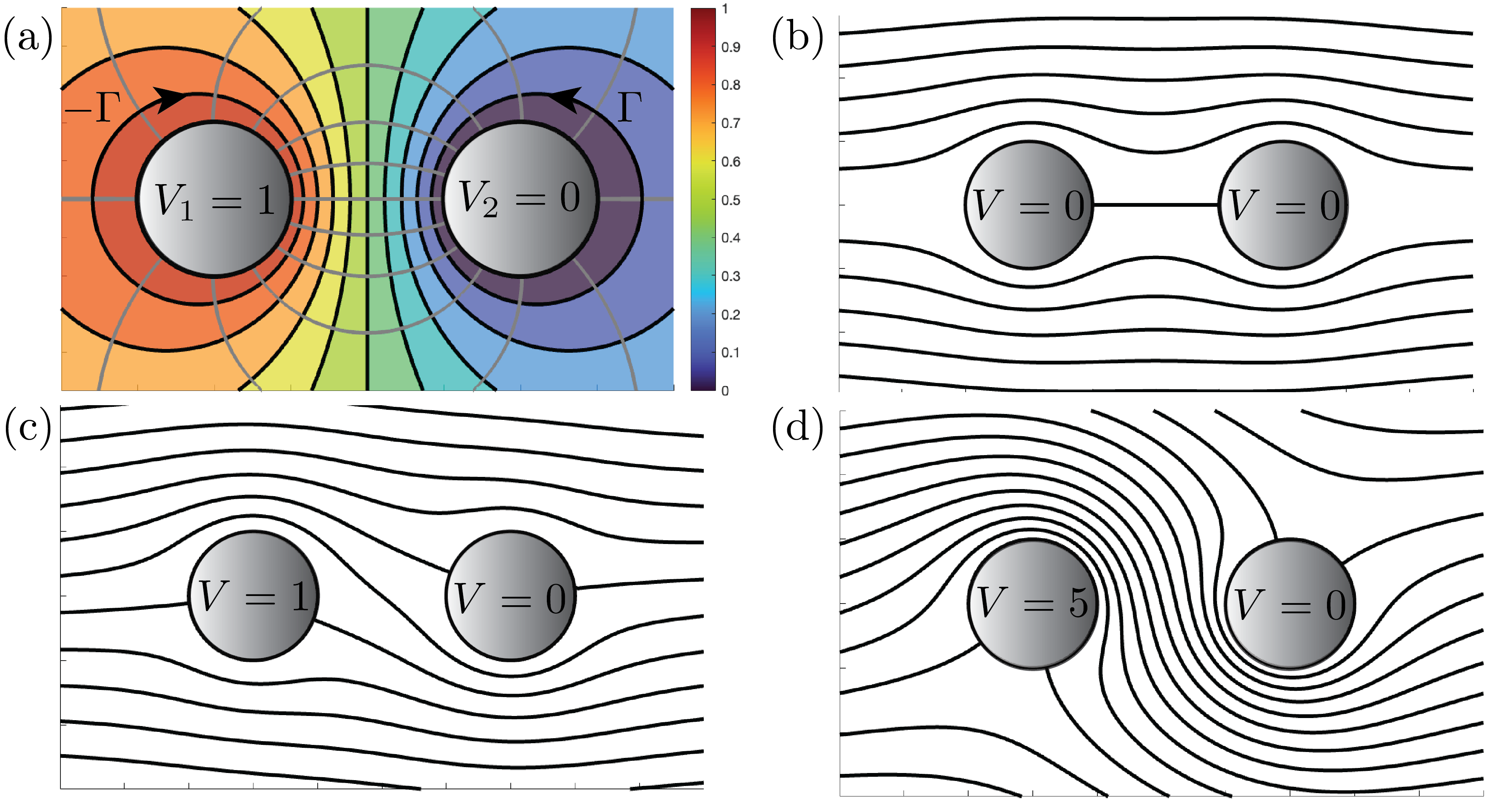}}
  \caption{(a) Visualization of the flow solution given in (\ref{eq:lorpot}). Grey lines represent electric field lines and the direction of current flow, $\boldsymbol{J}$. Black lines are streamlines of the fluid flow. A circulation of magnitude $\Gamma=|\sigma B_0 V_0|$ is induced around each body. (b) Black lines represent streamlines of a uniform flow past two conducting bodies. The given flow may be driven by either pressure or an external electric field since both solutions are identical. (c) Streamlines of a uniform flow past two cylinders when a potential difference $\Delta V =1$ is applied between the two cylinders, which drives electrical current and hence a circulation around each body. (d) Same plot as panel (c) with a larger applied voltage differential, and hence larger electrical current, between the cylinders. The larger current induces more circulation around each body as compared to panel (c).}
\label{fig:condcase}
\end{figure}

The particular map to the concentric annulus was already presented in (\ref{eq:map}). By direct manipulation, it may be shown that the map can be re-expressed as $z=(1-|a|^2)/(\zeta-a)-\overline{a}-d$, where $a=-d+\sqrt{d^2-1}$. Recasting the map in this form makes it clear that $\zeta=a$ is the pre-image of infinity. We now seek a complex potential which, in the physical domain, possesses zero circulation around each body and which tends to $W_{\mathrm{flow}}\sim U z$ as $|z|\rightarrow \infty$. 




The free-stream condition in the $\zeta$-plane becomes $W_{\mathrm{flow},\zeta}\sim U(1-|a|^2)/(\zeta-a)$ as $\zeta\rightarrow a$. The impermeability condition on rigid boundaries may be expressed as $\mathrm{Im}\left\{ W_{\mathrm{flow},\zeta}(\zeta) \right\}=\tilde{C}_j$ for $z\in \partial B_j$ for some set of constants $\{\tilde{C}_j\}$. A function which possesses all of the necessary properties, in the canonical domain, can be written concisely as follows,
\begin{subeqnarray}\label{eq:unif_nocirc}
    W_{\mathrm{flow},\zeta}(\zeta) & = & U\left(1-|a|^2\right)\phi_0(\zeta,a), \\[3pt]
    \phi_0(\zeta,a) & = & -\frac{1}{a}\mathcal{K}\left(\zeta,a\right)+\frac{1}{a}\mathcal{K}\left(\zeta,\frac{1}{\overline{a}}\right).
\end{subeqnarray}
where $\mathcal{K}\left(\zeta,a\right)=a\partial \log{\left(\omega(z,a)\right)}/\partial a$ and $\omega$ is the prime function as developed in \cite{crowdy2020solving}. The function $\phi_0$ has the two important properties: it has a simple pole with unit residue at $\zeta=a$, and it maps the circles of the canonical domain to slits parallel to the real axis, which possess a constant imaginary part (see \citet[pp. 83]{crowdy2020solving}). Hence, $\phi_0$ satisfies the impermeability condition on each body. The $\mathcal{K}$ functions in (\ref{eq:unif_nocirc}) possess a simple series representation in the doubly-connected case \citep[pp. 280]{crowdy2020solving}, which was used to generate the plot of flow streamlines given in figure \ref{fig:condcase}(b). Note that the solution in the physical domain is simply given by $W_{_{\mathrm{flow}},\zeta}{\left(\zeta(z)\right)}$, which is computed by substituting (\ref{eq:map}) into (\ref{eq:unif_nocirc}).


We note that if instead there were $N>2$ cylinders in the flow, the same solution (\ref{eq:unif_nocirc}) applies. However the definition of $\mathcal{K}$ changes. In such a case, a M{\"o}bius map converts the physical domain into a canonical domain comprising a unit disk with $N-1$ excised disks. The definition of $\mathcal{K}$ depends on the form of this canonical domain through $\omega$ and, in higher connectivities, it must be computed numerically by methods described in \cite{crowdy2007computing} and \citet[Chapter 14]{crowdy2020solving}.

\subsubsection{Magnetic Driving Induces Circulation}
In this section, we demonstrate how magnetic driving may be used to induce circulation into an otherwise circulation-free pressure driven flows such as that described in \S \ref{unifnocurr} and illustrated in figure \ref{fig:condcase}(b). The solution for a flow comprising a pressure-driven free stream in addition to magnetic driving can be obtained simply via superposition of the solutions from \S \ref{voltonly} and \S \ref{unifnocurr}.

In figure \ref{fig:condcase}(b), streamlines for the flow of a uniform stream in a Hele-Shaw cell are plotted using (\ref{eq:unif_nocirc}). Meanwhile, in figure \ref{fig:condcase}(a), the flow streamlines of the magnetic-driven flow from \S \ref{voltonly} are plotted. Figure \ref{fig:condcase}(c) then plots flow streamlines when a pressure-driven uniform stream is combined with a circulatory magnetic flow, of the type described in \S \ref{voltonly}.
Figure \ref{fig:condcase}(d) shows the streamlines when the intensity of magnetic flow is increased relative to  the situation in \ref{fig:condcase}(c); therein, the relative voltage between the conducting probes is increased by a factor of five which increases the magnitude of circulation around each body. The resulting flow streamlines in figure \ref{fig:condcase}(d) are successfully diverted between the cylinders.

Increasing the relative voltage between probes increases the magnitude of induced circulations, leading to a stronger diversion of the fluid flow from the uniform stream profile. Alternatively, the circulation can be enhanced for a fixed voltage differential by adjusting the geometry. As seen in (\ref{eq:elecpot}), the electrostatic capacitance of the two-cylinder system is $C=2\pi/\log{\left(1/\rho\right)}$. Since the capacitance is a conformal invariant, any configuration of conductors that can be conformally mapped to an annulus with inner radius $\rho$ and outer radius unity will induce the same circulation around each of the two conductors in the flow (for a fixed voltage difference). Thus, by changing the geometry in a conformally inequivalent manner, the induced circulation may thus be changed without modifying the applied voltage difference. See \cite{Leviconds} for a related discussion in the electrostatic context.

\subsubsection{General $N$-body Solutions Using Framework of Crowdy}\label{nbod}
Up to this point, we have examined the two-body problem to illustrate the physical mechanism for circulation generation. 
\begin{figure}
  \centerline{\includegraphics[width=0.95 \textwidth]{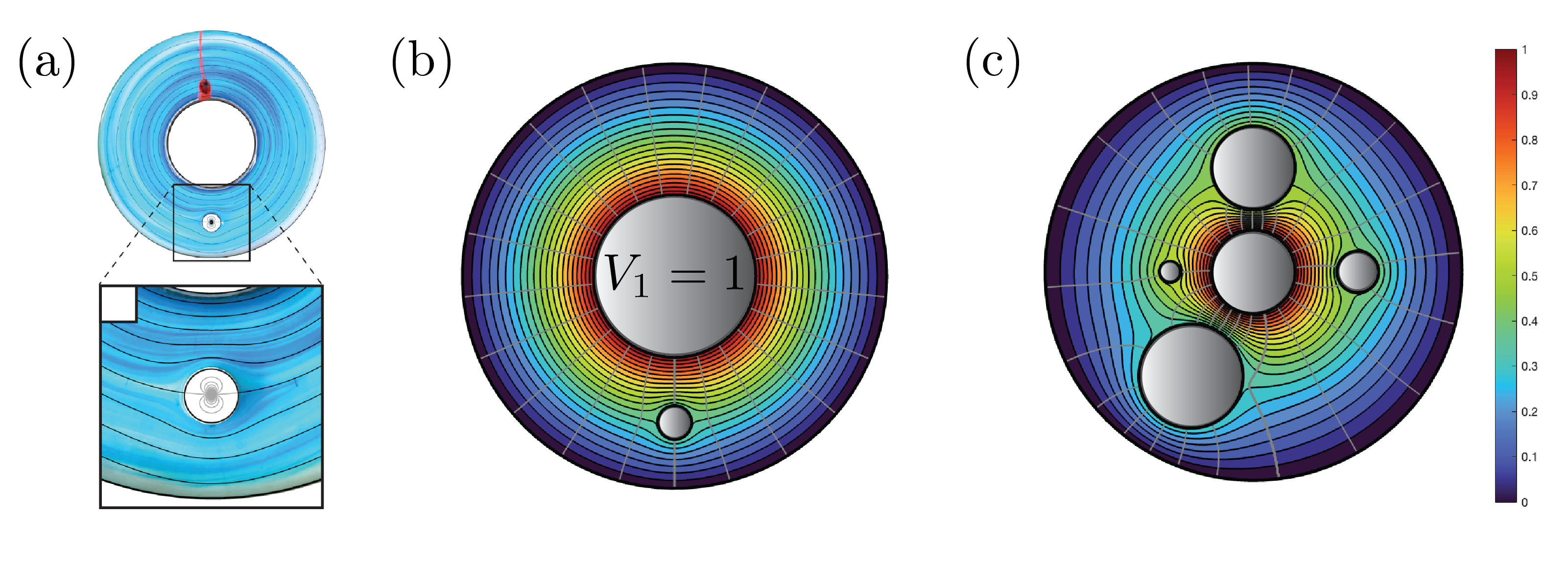}}
  \caption{(a) Experimental geometry taken directly from \cite{david2023magnetostokes}. (b) Exact solution, as obtained through the procedure of \S \ref{nbod}, for the experimental geometry from panel (a). Flow streamlines are depicted in grey. The voltage distribution is indicated in colour. (c) Another mathematical solution, as obtained through the procedure of \S \ref{nbod}. The central circle is held at $1\mathrm{V}$ while the outer circle is held at $0\mathrm{V}$. Other circles are taken to be floating conductors.}
\label{fig:experfig}
\end{figure}
Any two-body problem can be conformally mapped onto a concentric annulus having some inner radius $\rho$ and an outer radius $1$, where the value of $\rho$ depends on the geometry. In contrast to the Riemann mapping theorem — applicable for simply-connected domains — not all doubly-connected domains are necessarily conformally equivalent (they are equivalent only if $\rho$ happens to be the same). More generally, any $N$-body problems can be mapped to the interior of a unit circle with $N-1$ excised circles whose positions and radii are determined by the physical geometry. The concentric annulus is the special case of exactly one excised circle. To obtain the fluid flow around a number of physical obstacles, such as $N$ cylinders, it suffices to solve a problem in a conformally equivalent canonical domain, and then to map back to the physical domain of interest. Exact solutions for problems in the canonical domain are accessible through the framework of \citet{crowdy2020solving}. Note that to evaluate the solution in the physical domain, one must possess a conformal map to the physical domain of interest. 
In arbitrary geometries, the map between the physical and canonical domains may be difficult to attain analytically.

The procedure for solving the $N$-body problem is as follows. First, one must find a mapping from the physical domain of interest to a canonical domain comprising the unit circle with $N-1$ excised disks. \citet{crowdy2020solving} showed that this map can be written exactly in terms of the so-called prime function $\omega(z,a)$, for a number of physically relevant domains. For the case of the unbounded domain exterior to finitely many cylinders of arbitrary position and radius, a simple M{\"o}bius transformation brings the domain onto a canonical domain, as was described in \S \ref{unifnocurr}. Once the mapping to the canonical domain is found, the boundary value problem can be solved there instead. We now focus on solutions in canonical domains.

The solution in the canonical domain is amenable to the techniques developed by \citet{crowdy2020solving} and is expressible exactly in terms of the prime function. In relation to \S \ref{voltonly}, we seek a complex analytic function $W_E$ defined on a canonical domain which satisfies \ref{eq:Webvp}(a) on the unit circle and each excised disk. The geometry in figure \ref{fig:experfig}(c) represents $N_C=6$, so that there are five excised circles. In what follows we only consider only situations where all boundaries are perfectly conducting, so that $N_I=0$, and where external voltages are applied to each conductor. We also allow the possibility that some of the conductors are floating and not explicitly connected to a voltage source. We now describe how one may obtain solutions for the induced fluid flow in such geometries.

\citet[]{crowdy2020solving} introduced a special set of functions called integrals of the first kind, $\{v_j(z)\}$, defined in the canonical domain in terms of the prime function $\omega$, which will serve as the basis for the solutions obtained in the present section. \cite{crowdy2007computing} and \citet[Chapter 14]{crowdy2020solving} present efficient methods for computing the prime function. Codes for computing the prime function are readily available \citep{crowdy2016schottky}.
 
 There exists one such function, $v_j(z)$, for each of the excised circles so that $j\in \{1,2,...,N_C-1\}$, each possessing the following two important properties. First, $v_j(z)$ introduces a unit circulation around the $j^{\mathrm{th}}$ excised circle and exactly zero circulation around each of the other excised circles. Second, the imaginary part of $v_j(z)$ is constant on all other boundaries. It is thus clear that a linear superposition of the functions $\mathrm{i}v_j(z)$ is capable of meeting the boundary conditions in \ref{eq:Webvp}(a); 
the coefficients of the superposition are determined through the solution of a linear system of $N_C$ equations which is easily solved with the backslash operator in MATLAB. Details are presented in Appendix \ref{vjsecapp}. Once the electrostatic potential has been obtained, pre-multiplication by $-\mathrm{i}\sigma B_0$ gives the complex potential of the fluid flow as was described in \S \ref{compvar}.

 Solutions in two canonical domains, as obtained through the solution of a linear system of equations (see Appendix \ref{vjsecapp}), are presented in figures \ref{fig:experfig}(b) and \ref{fig:experfig}(c). Figure \ref{fig:exper}(a) shows a geometry that was explored experimentally by \cite{david2023magnetostokes}, wherein the authors were unable to attain an analytic solution. Figure \ref{fig:experfig}(b) shows the streamlines of the exact solution for the fluid flow in the same geometry as obtained through the method of the present section. The central cylinder is held at $V=1$ and the outer cylinder at $V=0$. The off-centre cylinder is taken to as a floating conductor, since it was not connected to voltage source in the experiments. As a consequence, there is no net circulation around the floating conductor. In figure \ref{fig:exper}(c), we plot flow streamlines in a domain of higher connectivity where $N_C=6$. The central cylinder is held at $V=1$ and the outer cylinder at $V=0$. The other conductors are chosen to be floating. Note that any combination of the conductors can be taken to be floating or possessing a prescribed voltage.

\subsection{Experiment}\label{exp}
Here we outline a simple experiment which realizes one of the exact solutions from \S \ref{nbod}. This experiment is meant to motivate and supplement the present theoretical paper but we note that it does not constitute a completely general experimental investigation of the magnetohydrodynamic Hele-Shaw cell. Future studies may focus on a more detailed experimental investigation of the system.

A Hele-Shaw cell was constructed from two thin circular sheets of transparent acrylic in a configuration similar to figure \ref{fig:schem}(a). Each transparent sheet had a diameter of $8\mathrm{cm}$ and a thickness of $1.5 \mathrm{mm}$. Two circles, with diameters of $d_1=7\mathrm{mm}$ and $d_2=10\mathrm{mm}$, were cut from each disk. Then, six small holes of diameter $1\mathrm{mm}$ were cut in the top acrylic sheet along the line passing through the two circles of radii $d_1$ and $d_2$, for the purposes of streamline visualization. A small drop of blue dye was place atop each such hole just before the experiment began. All cuts were made using a laser cutter.

\begin{figure} \centerline{\includegraphics[width=0.95 \textwidth]{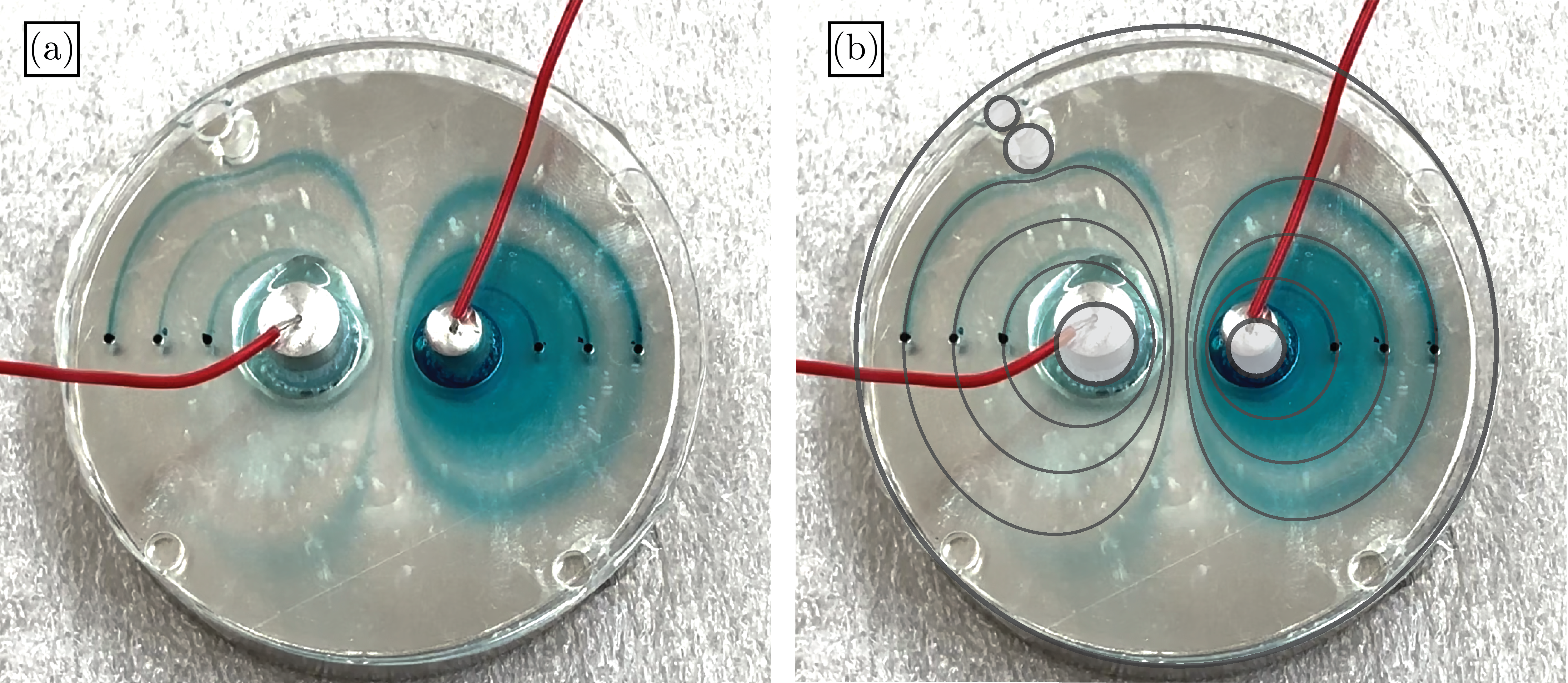}}
  \caption{(a) Top-down view of the Hele-Shaw cell filled with saltwater. The cell sits atop a permanant neodymium magnet. Two aluminum cylinders, of radii $7\mathrm{mm}$ and $10\mathrm{mm}$, completely penetrate the Hele-Shaw and are held at a voltage difference of $1\mathrm{V}$. (b) Same as panel (a) with mathematical solutions, as presented in \S \ref{nbod}, plotted atop the experimental streamlines. The bubbles in the top left of the Hele-Shaw cell are modelled as impermeable obstacles in the exact solution.}
\label{fig:exper}
\end{figure}

Note that four additional circular holes can be seen in seen in figure \ref{fig:exper}(a) near the boundary of each acrylic sheet; these holes serve no function in the present experiment. However, near the top left such hole, an unintentional bubble appeared when filling the cell, which will be addressed below.

The two sheets were then separated by spacers of thickness $h=0.7\mathrm{mm}$. The sheets were glued in place at each spacer, at several points along the boundary. Two metal cylinders, with diameters $d_1$ and $d_2$, were then fitted into the cell. The cylinders fit snugly into the bottom acrylic sheet in such a way that no glue was necessary to prevent leakage. In the upper acrylic sheet, the holes for the cylinders were made slightly larger than the cylinder diameters to allow gas to escape in the event of electrolysis.

The entire Hele-Shaw cell was then placed atop a DZ08-N52 Neodymium Disc Magnet, as ordered from K\&J Magnetics, with nominal surface field strength of $2340 \mathrm{G}$. The cell was then completely filled with saltwater using a needle. The saltwater was produced by simply adding salt packets to tap water. A small drop of blue dye was then placed atop each of the six $1\mathrm{mm}$ diameter holes just before the experiment began. A voltage difference of $1\mathrm{V}$ was then applied between the two cylinders. As the flow developed, the dye traced out streamlines as can be seen in figure \ref{fig:exper}(a).

Streamlines of exact solutions, as obtained though the procedure of \S \ref{nbod}, are plotted in figure \ref{fig:exper}(b). Only the six thoreotical streamlines which intersect the dye release points are plotted. The unintended bubble (top left) is treated as an impermeable boundary in the theoretical solution as is drawn in figure \ref{fig:exper}(b). There is a reasonable agreement between the experimental and theoretical streamlines.

Streamlines around the right-most cylinder are diffuse because the first few droplets of dye were dropped into place from too high and consequently spread beyond the small $1\mathrm{mm}$ opening. At the dye release points near the left cylinder, drops were added more carefully by gently touching a droplet onto the small opening, giving rise to more defined streamlines.


Note that the solution in figure \ref{fig:exper}(b) treats the bubbles as floating electrical conductors since the framework of \S \ref{nbod} only allows for conducting obstacles. In reality, a bubble is better approximated by an insulator. However, since there are only two electrodes with fixed applied voltages in the experimental geometry (figure \ref{fig:exper}(a)), the flow streamlines are unaffected by this assumption. That is to say, the streamlines in figure \ref{fig:exper}(b) would be identical to those found if bubbles were treated as perfect insulators. However, the magnitude of the induced flow velocities does in fact depend on the type of boundary condition imposed: the induced circulation is higher in the presence of floating conductors compared to insulators. Generally, when there are multiple applied voltages, one must apply the appropriate insulating boundary condition on insulating obstacles in order to obtain accurate flow streamlines. Even in the case of figure \ref{fig:exper}(a), the proper insulating boundary condition needs to be applied in order to obtain accurate values for the velocity magnitudes at each point in the flow. A numerical scheme which allows as well for the presence of insulating obstacles, which thus overcomes the limitations of \S \ref{nbod}, is presented in \S \ref{num}.

\section{Series Solutions}\label{num}
When no conformal mapping is known between the physical domain of interest and a canonical domain, or if insulating bodies exist in the flow, solutions are not attainable by the method presented in \S \ref{nbod}. Moreover, even when solutions can be written explicitly in terms of the prime function, as in \S \ref{nbod}, the numerical computation of the prime function can be expensive and actually involves the numerical solution of a different boundary value problem \citep{crowdy2016schottky}.

All of these facts lead us to seek out accurate and efficient numerical solutions for the fluid flow, in general settings. In this section, we demonstrate how series solutions can be adapted to our problem to solve for the fluid flow with many digits of accuracy in just a few seconds of computing time on a standard laptop \citep{trefethen2018series}. The procedure is described as follows.

The complex potential described by (\ref{eq:Webvp}) is first expressed as a sum of Laurent series centered inside each body (exterior to the flow domain). The Laurent series are then truncated and their coefficients are determined through a least squares problem which enforces the conditions (\ref{eq:Webvp}) at a finite number of points on each boundary.

The Vandermonde matrix in the least squares problem becomes exponentially ill-conditioned in the degree of approximation. As a result, the numerical solution eventually saturates with an increasing degree of approximation.  
\cite{brubeck2021vandermonde} showed that by performing Arnoldi orthogonalization, one can improve the condition number and thus achieve more digits of accuracy in the numerical solution. In some least squares problems, the difference in accuracy between solutions obtained with and without Arnoldi can be quite dramatic, even reaching ten digits \citep[see figure 3.1]{brubeck2021vandermonde}. Note that for the examples considered in the present section, the Arnoldi orthogonalization is not essential for attaining quite accurate solutions. In more general geometries, the Arnoldi procedure may be necessary to achieve high accuracy solutions.



The numerical method described herein is similar to that presented by \cite{baddoo2020lightning}, but with modified boundary conditions and an extension to multiply-connected domains. Note that when sharp corners are present in the solution domain, strong singularities in the solutions emerge, and rapidly converging lightning solvers can be applied \citep{gopal2019solving}. Lightning solvers supplement the Laurent series representation with a set of poles clustered near sharp corners to approximate singularities. In the present work, we focus on smooth bodies which do not require the placement of such poles. However, the procedure for including such poles is rather straightforward \citep{baddoo2020lightning,gopal2019solving}.


\subsection{Numerical Solution of Electrostatic Problem}\label{numelec}
We seek the electrostatic potential in the unbounded domian exterior to $N =N_I+N_C$ distinct bodies, with $N_C$ conducting surfaces $\{\partial B_j\}$ held generally at different voltages $V_j$, and $N_I$ insulating surfaces $\{\partial D_j\}$ obeying zero Neumann conditions $\nabla V_j\boldsymbol{\cdot}\boldsymbol{n}=0$.
For notational convenience, let $z_j$ be a point interior to the $j^{\mathrm{th}}$ conductor for $j\in\{1,...,N_C\}$ and the interior of the $\left(j-N_C\right)^{\mathrm{th}}$ insulator for $j\in\{N_C+1,...,N_C+N_I\}$.

The electrical potential then takes the form (\ref{eq:WEsourcebehav}), where $Q_i$ are undetermined constants. The solution can be written generally as a sum of a Laurent series and its logarithm terms as
\begin{subeqnarray}\label{eq:numelformlaur}
    W_E(z) & = & a_0 + \sum_{j=1}^{N_C} Q_j \log{\left(z-z_j\right)}+\sum_{j=1}^N\sum_{k=1}^{\infty} \frac{a_k^{(j)}}{\left(z-z_j\right)^k},  \\[3pt]
    \sum_{j=1}^{N_C} Q_j & = & 0, \\[3pt]
    \mathrm{Re}\left\{W_E\right\} & = & V_j,\;z\in \partial B_j,\;j\in \{1,2,...,N_C\}, \\[3pt]
    \mathrm{Im}\left\{W_E\right\} & = & C_j,\;z\in \partial D_j,\;j\in \{1,2,...,N_I\},
\end{subeqnarray}
where $C_j$ and $Q_j$ are apriori unknown real constants. Note that no logarithm terms are centered inside of electrical insulators. The condition (\ref{eq:numelformlaur}b) is required to ensure that the potential is regular at infinity. 

To convert (\ref{eq:numelformlaur}) into a least-squares fitting problem, we must truncate each Laurent series. Let the expansion centered around the $j^{\mathrm{th}}$ body be truncated at $N_L^{j}$ terms. Combining (\ref{eq:numelformlaur}a) and (\ref{eq:numelformlaur}b), the form of our approximation becomes
\begin{equation}\label{eq:numerapprox}
    W_E(z)  \approx  a_0 + \sum_{j=2}^{N_C} Q_j \log{\left(\frac{z-z_j}{z-z_1}\right)}+\sum_{j=1}^N\sum_{k=1}^{N_L^{j}} \frac{a_k^{(j)}}{\left(z-z_j\right)^k},
\end{equation}
where we have implemented (\ref{eq:numelformlaur}b) by setting $Q_1=-\sum_{j=2}^{N_C}Q_j$.
Through this implementation, we guarantee that (\ref{eq:numelformlaur}b) is satisfied exactly so that $W_E(z)$ is finite as $|z|\rightarrow\infty$. If (\ref{eq:numelformlaur}b) were instead implemented as a constraint in the least-squares problem, then (\ref{eq:numelformlaur}b) would only hold approximately and $W_E(z)$ would not be guaranteed to be finite as $|z|\rightarrow \infty$. We next choose $N_b^{j}$ sample points on the boundary of the $j^{\mathrm{th}}$ body and fit $W_E$ to the appropriate boundary conditions, (\ref{eq:numelformlaur}c)-(\ref{eq:numelformlaur}d). 

We now formulate the least squares problem. Let $z_j^{(k)}$ denote the $j^{\mathrm{th}}$ sample point on the surface of the $k^{\mathrm{th}}$ body, where $k\in\{1,\cdots,N_I+N_C\}$ and $j\in\{1,\cdots,N_b^k\}$. In order to construct the Vandermonde matrix, we first define the vector
\begin{equation}
\boldsymbol{V}_j^{(k)} = \left[
\begin{array}{ccc|c|ccc}
  (z_j^{(1)}-z_1)^{-1}  & \cdots & (z_j^{(1)}-z_1)^{-N_L^{1}} & \cdots &  (z_j^{(1)}-z_N)^{-1}  &  \cdots & (z_j^{(1)}-z_N)^{-N_L^{N}}
\end{array}  \right] .
\label{defQc}
\end{equation}
We next define a vector of logarithm term evaluations corresponding to the $j^{\mathrm{th}}$ sample point on the $k^{\mathrm{th}}$ boundary as follows,
\begin{equation}
\boldsymbol{L}_j^{(k)} = \left[
\begin{array}{ccc}
  \log{\left(\frac{z_j-z_2}{z_j-z_1} \right)} & ... & \log{\left(\frac{z_j-z_{N_C}}{z_j-z_1} \right)}\\
\end{array}  \right] .
\label{defQc}
\end{equation}
We also define the Laurent series coefficient vector as
\begin{equation}
\boldsymbol{a} = \left[
\begin{array}{ccc|c|ccc}
  a_1^{(1)} & ... & a_{N_L^1}^{(1)} & ... & a_1^{(N)} & ... & a_{N_L^N}^{(N)}
\end{array}  \right] .
\label{defQc}
\end{equation}

Now let $\boldsymbol{V}^{(k)}$ be the matrix of containing all sample points on the $k^{\mathrm{th}}$ body defined by 
\begin{equation}
\boldsymbol{V}^{(k)} = \left[
\begin{array}{c}
  \boldsymbol{V}_1^{(k)}\\
  \boldsymbol{V}_2^{(k)}\\
  ...\\
  \boldsymbol{V}_{N_{b}^{k}}^{(k)}
\end{array}  \right].
\label{defQc}
\end{equation}
Similarly, we let $\boldsymbol{L}^{(k)}$ define the matrix of logarithm evaluations on the $k^{\mathrm{th}}$ body,
\begin{equation}
\boldsymbol{L}^{(k)} = \left[
\begin{array}{c}
  \boldsymbol{L}_1^{(k)}\\
  \boldsymbol{L}_2^{(k)}\\
  ...\\
  \boldsymbol{L}_{N_{b}^{k}}^{(k)}
\end{array}  \right].
\label{defQc}
\end{equation}
Then, the least squares problem for the undetermined coefficients can be expressed in terms of the matrix $\boldsymbol{A}$ defined by
\begin{equation}
\boldsymbol{A} = \left[
\begin{array}{cccc|c|cccc}
  \boldsymbol{0} & \boldsymbol{1} &-\mathrm{Im}\left\{\boldsymbol{V}^{(1)}\right\}  & \mathrm{Re}\left\{\boldsymbol{V}^{(1)}\right\} & \mathrm{Re}\left\{\boldsymbol{L}^{(1)}\right\} & \boldsymbol{0} & \boldsymbol{0} & ... & \boldsymbol{0}\\
  \boldsymbol{0} & \boldsymbol{1} &\vdots  & \vdots & \vdots & \boldsymbol{0} & \boldsymbol{0} & ... & \boldsymbol{0}\\
  \boldsymbol{0} & \boldsymbol{1} &-\mathrm{Im}\left\{\boldsymbol{V}^{(N_C)}\right\}  & \mathrm{Re}\left\{\boldsymbol{V}^{(N_C)}\right\}& \mathrm{Re}\left\{\boldsymbol{L}^{(N_C)}\right\} & \boldsymbol{0} & \boldsymbol{0} & ... & \boldsymbol{0}\\
  \boldsymbol{1} & \boldsymbol{0} &\mathrm{Re}\left\{\boldsymbol{V}^{(N_C+1)}\right\}  & \mathrm{Im}\left\{\boldsymbol{V}^{(N_C+1)}\right\}& \mathrm{Im}\left\{\boldsymbol{L}^{(N_C+1)}\right\} & -\boldsymbol{1} & \boldsymbol{0} & ...& \boldsymbol{0}\\
  \boldsymbol{1} & \boldsymbol{0} &\mathrm{Re}\left\{\boldsymbol{V}^{(N_C+2)}\right\}  & \mathrm{Im}\left\{\boldsymbol{V}^{(N_C+2)}\right\}& \mathrm{Im}\left\{\boldsymbol{L}^{(N_C+2)}\right\} & \boldsymbol{0} & -\boldsymbol{1} & \boldsymbol{0} & ...\\
  \boldsymbol{1} & \boldsymbol{0} &...  & ...& ... & ... & ... & ... & ...\\
  \boldsymbol{1} & \boldsymbol{0} &\mathrm{Re}\left\{\boldsymbol{V}^{(N)}\right\}  & \mathrm{Im}\left\{\boldsymbol{V}^{(N)}\right\}& \mathrm{Im}\left\{\boldsymbol{L}^{(N)}\right\} & \boldsymbol{0} & \boldsymbol{0} & \boldsymbol{0}& -\boldsymbol{1}\\
\end{array}  \right],
\label{defQc}
\end{equation}
along with the accompanying coefficient vector $\boldsymbol{c}$ defined by
\begin{equation}
\boldsymbol{c} = \left[
\begin{array}{ccccccccc}
  \mathrm{Im}\left\{a_0\right\} & \mathrm{Re}\left\{a_0\right\} &\mathrm{Im}\left\{\boldsymbol{a}\right\} & \mathrm{Re}\left\{\boldsymbol{a}\right\} & Q_2 & ... & Q_{N_C} & C_1 & ... C_{N_I}
\end{array}  \right].
\label{defQc}
\end{equation}
Note that the vectors $\boldsymbol{1}$ and $\boldsymbol{0}$ in the definition of $\boldsymbol{A}$ represent vectors of ones and zeros of the appropriate length; for example, the vector $\boldsymbol{1}$ in the top row has a length of $N_b^1$ while the $\boldsymbol{1}$ in the bottom row has a length of $N_b^N$. The load vector, having a length equal to the total number of boundary sample points, $\sum_{j=1}^N N_b^j$, then takes the form
\begin{equation}\label{eq:finalmat}
\boldsymbol{f} = \left[
\begin{array}{ccc|c|ccc|ccc}
  V_1 & \cdots & V_1 & \cdots &V_{N_C} & \cdots & V_{N_C} & 0 & \cdots & 0
\end{array}  \right],
\label{defQc}
\end{equation}
with which the approximate form of (\ref{eq:numelformlaur}) is expressed as the following matrix equation,
\begin{equation}\label{eq:LSelec}
\boldsymbol{A}\boldsymbol{c}^T \approx \boldsymbol{f}^T.
\end{equation}
Equation (\ref{eq:LSelec}) can be solved with the backslash operator in MATLAB for the coefficient vector $\boldsymbol{c}$, from which the complex potential can be readily evaluated using (\ref{eq:numerapprox}).
Generally, the solution accuracy can be greatly improved by using Arnoldi orthogonalization on the Vandermonde part of the matrix $\boldsymbol{A}_{\mathrm{flow}}$. We omit the details of the Vandermonde orthogonalization procedure for brevity. Details of the procedure may be found in \cite{brubeck2021vandermonde} and an application to potential flow is presented by \cite{baddoo2020lightning}. Arnoldi orthogonalization may be applied to the least squares problem (\ref{eq:LSelec}) through simple modifications to code given by \cite{brubeck2021vandermonde}.

\begin{figure}
  \centerline{\includegraphics[width=0.95 \textwidth]{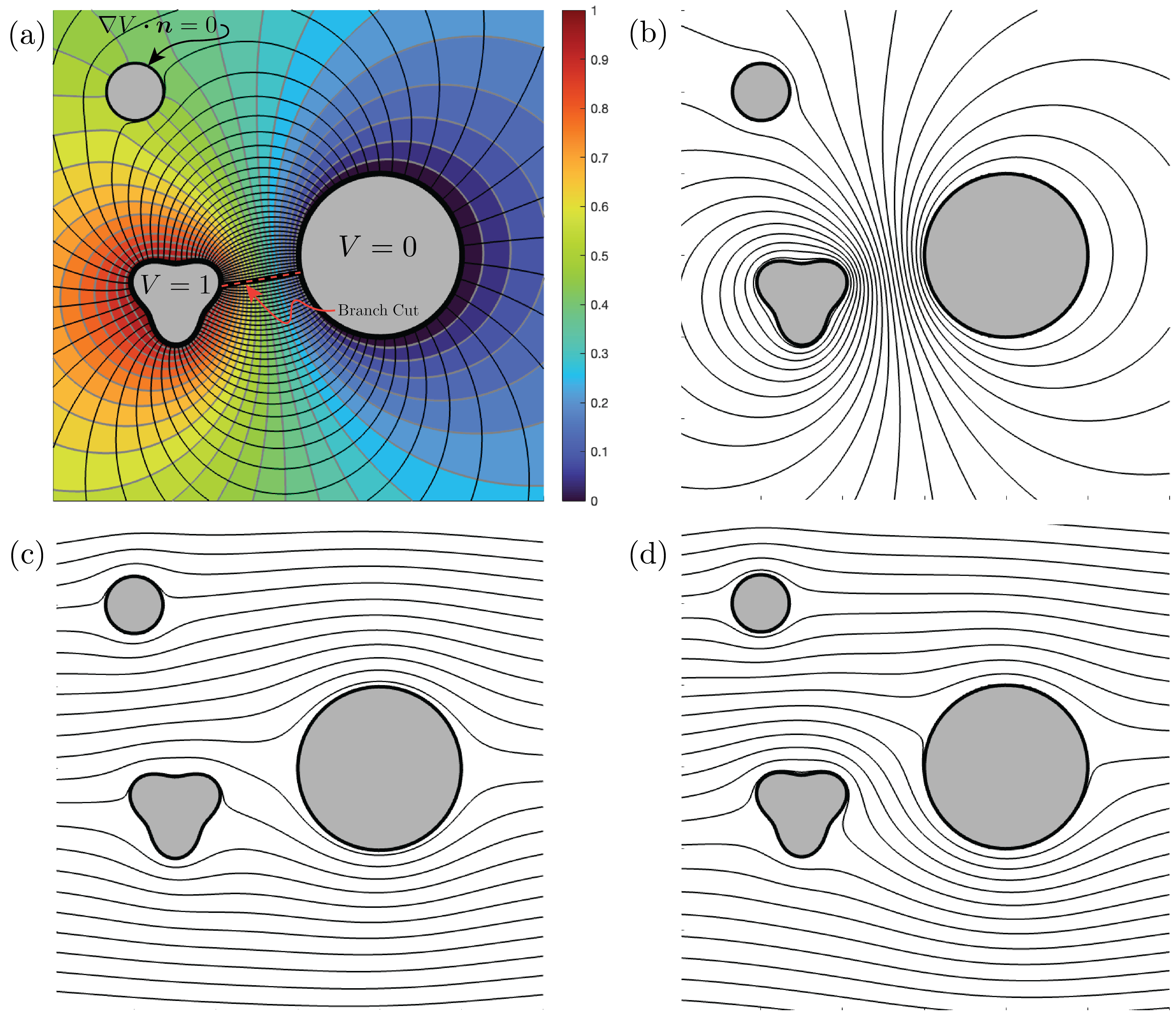}}
  \caption{(a) Numerical solution of the electrostatic problem involving two perfect conductors held at voltages $V=1$ and $V=0$ along with a perfect insulator, obtained though the approach outlined in \S \ref{numelec}. The charge magnitude on each conducting surface, $Q$, is obtained through the solution of the least squares problem.(b) The fluid flow induced by the voltage distribution of (a), in the presence of an out-of-plane magnetic field. The flow is computed through the procedure outlined in \S \ref{numfluid}. The magnitude of the circulation around each conductor is specified through the solution presented in (a), $\Gamma =\sigma B_0 Q$. (c) A free-stream fluid flow past the same obstacles in a Hele-Shaw cell in the absence of magnetic effects (no circulation). (d) The free stream flow from (c) combined with magnetically induced circulation due to the electrical configuration in part (a). The fluid flow depicted in (d) is essentially a superposition of the flows depicted in (b) and (c).}
\label{fig:numericalstuff}
\end{figure}

\subsection{Finding the Fluid Flow}\label{numfluid}
If all of the bodies in the flow are perfect electrical conductors, then the complex potential describing the fluid flow is obtained by pre-multiplying the electrical potential from \S \ref{numelec} by $-\mathrm{i}\sigma B_0$, which follows from  (\ref{eq:FDEF}a) and (\ref{eq:floweq_comp}), since in this case $W_P=0$ as was discussed in \S \ref{compvar}. 

When one or more of the obstacles in the flow is an electrical insulator, the potential $-\mathrm{i}\sigma B_0 W_E$ violates the impermeability condition on electrical insulators. This is illustrated in figure (\ref{fig:numericalstuff}a), where the flow streamlines of $-\mathrm{i}\sigma B_0 W_E$ (shown in grey) clearly penetrate the insulating body. In cases where insulating bodies are present, the fluid flow must thus be obtained in two steps. First, the electrostatic problem must be solved to specify the induced circulation in the fluid flow around each body as in (\ref{eq:Wfform}). Then, the fluid flow boundary value problem, subject to impermeable boundary conditions on each obstacle, must be solved. When electrical insulators are present in the flow, a non-zero pressure arises whose role is to enforce the impermeability condition.

The complex potential for the fluid flow problem can be written as follows,
\begin{subeqnarray}\label{eq:numelformlaurflow}
    W_{\mathrm{flow}}(z) & = & b_0 + \sum_{j\in I_C} \mathrm{i}\sigma B_0 Q_j \log{\left(z-z_i\right)}+\sum_{j=1}^N\sum_{k=1}^{\infty} \frac{b_k^{(j)}}{\left(z-z_j\right)^k}  +\overline{U}z, \\[3pt]
    \mathrm{Im}\left\{W_E\right\} & = & \psi_j,\;z\in \partial B_j,\;j\in\{1,2,...,N\},
\end{subeqnarray}
where $\{b_k^{(j)}\}$ and $\{\psi_j\}$ are unknown coefficients, $U$ is a background pressure-driven flow, and $Q_j$ are the known charges from the electrostatic solution of \S \ref{numelec}. Taking $U=0$ gives the situation of a flow driven exclusively by magnetic effects. Note that we can immediately take $\mathrm{Re}\left\{b_0\right\}=0$, since the boundary value problem does not place any restriction on the real part of this constant.

The potential can be approximated by once again truncating the Laurent series centered in each body,
\begin{subeqnarray}\label{eq:numelformlaurflow}
    W_{\mathrm{flow}}(z) & \approx & b_0 + \sum_{j\in I_C} \mathrm{i}\sigma B_0 Q_j \log{\left(z-z_i\right)}+\sum_{j=1}^N\sum_{k=1}^{N_L^{(j)}} \frac{b_k^{(j)}}{\left(z-z_j\right)^k}  +\overline{U}z,
\end{subeqnarray}
which is still subject to (\ref{eq:numelformlaurflow}b).

The sets of coefficients, $b_k^{j}$ and $\psi_k$, can be obtained in a manner similar to \S \ref{numelec}. In this case, there are no undetermined coefficients associated with logarithm terms.
Compared to the electrostatic problem of \S \ref{numelec}, the matrix $\boldsymbol{A}$ is changed in two ways in the fluid flow problem. Firstly, logarithm terms appear in $\boldsymbol{f}$ instead of in $\boldsymbol{A}_{\mathrm{flow}}$. Second, since all bodies possess an undetermined stream function value, the right partition of $\boldsymbol{A}$ is changed accordingly. In the flow problem, the flow matrix $\boldsymbol{A}$ becomes
\begin{equation}
\boldsymbol{A}_{\mathrm{flow}} = \left[
\begin{array}{c|cc|cccc}
  \boldsymbol{1}   &\mathrm{Re}\left\{\boldsymbol{V}^{(1)}\right\}  & \mathrm{Im}\left\{\boldsymbol{V}^{(1)}\right\} & -\boldsymbol{1} & \boldsymbol{0} & \boldsymbol{0} & \boldsymbol{0} \\
  \boldsymbol{1} & \vdots  & \vdots & \boldsymbol{0} & -\boldsymbol{1} & \boldsymbol{0} & \vdots\\
  \boldsymbol{1} &\mathrm{Re}\left\{\boldsymbol{V}^{(N_C)}\right\}  & \mathrm{Im}\left\{\boldsymbol{V}^{(N_C)}\right\} & \vdots & \boldsymbol{0} & -\boldsymbol{1} & \ddots\\
  \boldsymbol{1} &\mathrm{Re}\left\{\boldsymbol{V}^{(N_C+1)}\right\}  & \mathrm{Im}\left\{\boldsymbol{V}^{(N_C+1)}\right\} & \vdots & \vdots & \boldsymbol{0}& \ddots\\
  \boldsymbol{1} &...  & ... & \vdots & \vdots & \vdots & \ddots\\
  \boldsymbol{1} &\mathrm{Re}\left\{\boldsymbol{V}^{(N)}\right\}  & \mathrm{Im}\left\{\boldsymbol{V}^{(N)}\right\} & \boldsymbol{0} & \boldsymbol{0} & \boldsymbol{0} & ...\\
\end{array}  \right],
\label{defQc}
\end{equation}
where $\boldsymbol{1}$ and $\boldsymbol{0}$ represent vectors of ones and zeros of the appropriate length; for example, the vector $\boldsymbol{1}$ in the top row has a length of $N_b^1$ while the $\boldsymbol{1}$ in the bottom row has a length of $N_b^N$. The corresponding coefficient vector becomes
\begin{equation}
\boldsymbol{c}_{\mathrm{flow}} = \left[
\begin{array}{cccccc}
  \mathrm{Im}\left\{a_0\right\} &\mathrm{Im}\left\{\boldsymbol{a}\right\} & \mathrm{Re}\left\{\boldsymbol{a}\right\} & \psi_1 & ...&  \psi_{N}
\end{array}  \right],
\label{defQc}
\end{equation}
The length of the new load vector $\boldsymbol{f}_{\mathrm{flow}}$
is the total number of sample points on all boundaries, $\sum_{k=1}^N N_b^{k}$ components. For the boundary points on the $k^{\mathrm{th}}$ body, the corresponding section of $\boldsymbol{f}_{\mathrm{flow}}$ is given by a row vector $\boldsymbol{f}^{(k)}$ whose $n^{\mathrm{th}}$ component is given by
\begin{equation}
f^{(k)}_n=\mathrm{Im}\left\{-\sum_{j=1}^{N_C} \mathrm{i}\sigma B_0 Q_j\log(z^{(k)}_n-z_j)-\overline{U}z^{(k)}_n\right\}.
\end{equation}
The new load vector is then given by concatenating the vectors corresponding to each boundary so that $\boldsymbol{f}_{\mathrm{flow}}=\left[
\begin{array}{ccc}
  \boldsymbol{f}^{(1)}&\cdots& \boldsymbol{f}^{(N)}
\end{array}  \right]$. The boundary value problem is then reduced to the following least squares problem,
\begin{equation}\label{eq:LSflow}
\boldsymbol{A}_{\mathrm{flow}}\boldsymbol{c}_{\mathrm{flow}}^T \approx \boldsymbol{f}_{\mathrm{flow}}^T.
\end{equation}
Once again, we note that the solution accuracy can be improved by using Arnoldi orthogonalization on the Vandermonde part of the matrix $\boldsymbol{A}_{\mathrm{flow}}$ when solving (\ref{eq:LSflow}). Figure \ref{fig:numericalstuff}(b) shows the fluid flow induced by the voltage configuration illustrated in figure \ref{fig:numericalstuff}(a). 

By checking the deviation of the imaginary part of the $W_{\mathrm{flow}}$ from a constant, we can attain an estimate for the error in $\mathrm{Im}\left\{W_{\mathrm{flow}}\right\}$ over the entire domain. Checking the boundary convergence on a set that is 16 times as dense as the sample points, we find an accuracy of ten digits on the circular boundaries when using 40 Laurent terms in each body and 200 uniformly sample points per body. The accuracy on the trefoil shaped boundary in figure \ref{fig:numericalstuff} is limited to a more modest seven digits owing to the finger-like geometry. In more extreme finger-like geometries, one should consider the placement of poles in addition to the Laurent series used in the present paper, to account for the well-known crowding phenomena \cite{gopal2019representation}. For work exploiting simple pole placement to achieve accurate solutions in the case of sharp corners and highly curved geometries, see the works of \cite{gopal2019solving}, \cite{costa2021aaa}, and \cite{yidan2024}. Note that \cite{baddoo2020lightning} applied such methods to the potential flow problem past a simply-connected body with a corner and he demonstrated rapid convergence when simple poles are exponentially clustered near corners.

\section{Discussion and Conclusion}
We have analysed the flow of an uncharged ohmic fluid inside a magnetically driven Hele-Shaw cell, at low Hartmann numbers. The problem was first cast the into a complex variables framework. Within this framework, we elucidated — both physically and mathematically — the mechanism by which an external magnetic field may convert electric current into a tunable fluid flow circulation. Whereas pressure-driven Hele-Shaw flows exhibit identically zero circulation around any closed contour, the circulation in magnetically driven Hele-Shaw flows need not vanish. Moreover, we have demonstrated that by changing the voltage difference applied between conducting probes in the fluid, as well as the probe geometries, one can control the induced fluid flow circulation.

Within our complex variables framework, we presented mathematical solutions, in terms of the prime function, for a class of geometries involving circular conductors; one such solution describes an experimental geometry of \cite{david2023magnetostokes} as is illustrated in figure \ref{fig:condcase}. Notably, the prime function has a series representation in doubly-connected geometries so that solutions become exact \citep{baddoo2020exact}. We note that more exact solutions in periodic domains are also available through the theoretical developments of \cite{baddoo2021calculus}; however, such results were not explored in the present paper.

We subsequently noted two limitations of the aforementioned mathematical solutions. Firstly, they are only applicable in canonical circular domains. To obtain a solution in non-circular geometries, a conformal map between a canonical domain and the physical domain of interest must be known. As such, the class of solutions that may be written explicitly in terms of the prime function is limited. Second, the exact solutions do not apply when any of the obstacles in the flow is electrically insulating. Moreover, it should be noted that the prime function must also be computed numerically in domains with a connectivity greater than two.

Because of these limitations, we proceeded to present a numerical scheme capable of finding the fluid flow in more general geometries and in situations where obstacles in the flow may be either insulating or conducting. We outlined a procedure for obtaining accurate Laurent series solutions for the flow in a manner similar to that described by \cite{trefethen2018series}.

Approximate solutions must be obtained in two steps. First, one must obtain a series solution to the relevant electrostatic problem. The electrostatic field $\boldsymbol{E}$ is then converted to an electric current via Ohm's Law, $\boldsymbol{J}=\sigma \boldsymbol{E}$. The net current $I_i$ leaving each conducting body $\partial B_i$, along with the external magnetic field strength $B_0$, dictates the circulation around the body in the ensuing fluid flow according to $\Gamma_i = B_0 I_i$. 
Both geometry and applied voltage magnitudes affect the magnitude of induced circulations.
The key result of the the first step is that the electrostatic problem specifies the fluid flow circulation around each conducting body. With the circulation around each body fixed, the potential flow problem is fully posed. The second step of the solution is to find a series solution for the potential flow subject to impermeable boundary conditions on each obstacle and with the circulations around each obstacle imposed by the electrostatic problem. A uniform stream, or other background flows, can be added at this stage by superposition.

Note that in the magnetohydrodynamic Hele-Shaw cell, the circulation around each body in the flow is specified. In the absence of magnetic fields, the velocity potential in a Hele-Shaw cell is proportional to the pressure and is therefore single-valued, corresponding to a flow with identically zero circulation. Meanwhile, when subjected to magnetic forcing, 
the Lorentz force generates a well-defined and calculable amount of circulation around each body. This situation is in contrast with potential aerofoil theory, where the circulation around each body is unknown. Since the velocity potential in aerofoil theory is not related to a physical quantity (such as pressure), no physical constraint enforces the circulation value around each body. In some special cases, such as for aerofoils with a single sharp corner, Kutta's condition specifies a unique circulation to de-singularize the velocity field. Recent developments by \cite{gonzalez2022variational} conjecture a new criterion to replace the Kutta condition in more general aerofoil geometries.

Magnetohydrodynamic flow control, as explored in the present study, has many potentially interesting applications in, for example, microfluidic devices. The voltage difference applied between probes, in the magnetically driven flow, may be actively controlled during an experiment to divert flow along desired paths. Probe geometries may also be designed to achieve desired flows. More engineering applications that might be be explored in future studies are outlined by \cite{bau2022applications}. Future theoretical work might examine the possibility of bubble manipulation using magnetically-driven flows in a Hele-Shaw cell \citep{booth2023circular}. In realizing such applications, the principles and solution methods of the present paper may be useful for quick and iterative design. Probe geometries and voltages can be adjusted very simply within the framework of \S \ref{num} making it and attractive tool for iterative design.

\textbf{Acknowledgements.} I would like to thank Cy David for his interesting presentation at the APS DFD conference in Washington, DC which served to inspire this investigation. I am also indebted to Peter Baddoo for introducing me to the wonderful mathematical playground of conformal maps and complex variables, before his tragic passing in February of 2023. I am certain that this work would not have been possible without the generous mentorship that Peter provided during his time as an instructor at MIT. I would also like to thank Darren Crowdy, Keaton Burns, and Bauyrzhan Primkulov for valuable discussions. Finally, I would like to thank Bauyrzhan Primkulov for helping with the experiment presented in \S \ref{exp}.

\paragraph{\textbf{Funding.} The author was funded with a MathWorks Fellowship during this work. During the completion of this paper, he was also supported by a Chateaubriand fellowship, and hosted at ESPCI, Paris.}

\paragraph{\textbf{Declaration of Interests.} } The author reports no conflict of interest.

\appendix
\section{Solution for Bodies at Different Potentials}\label{Diff_vels}
\subsection{The Dirichlet Problem in Canonical Domains}
Consider the $N$-body Dirichlet problem for the electrical potential. In particular, consider the canonical geometry in figure \ref{fig:experfig}(c) where excised circles of radius $q_j$ are centered at positions $\delta_j$. Each circle is held at some constant specified potential $V_j$. We seek the potential in the interstitial region between excised circles. Following \S \ref{compvar}, we represent $V_j$ as the real part of an analytic function $W_E$.

In \S \ref{voltonly}, the exact complex potential for the two-body problem was expressible in terms of the simple logarithm, (\ref{eq:elecpot}). As we will show, the potential $W_E$ in the canonical $N-$body domain, is expressible in terms of the prime function.

\subsection{Prime Function Machinery: Integral of the First Kind, $v_j(z)$}\label{vjsecapp}
We will present the essential components of the theory for the reader. For a more complete treatment, see \cite{crowdy2020solving}. \cite{crowdy2020solving} introduces the functions $v_j(z)$, defined in the canonical domain, which possess two important properties: 1) the imaginary part of $v_j(z)$ is constant on all circular boundaries; and 2) $v_j(z)$ possesses unit circulation around $\partial B_j$ and zero-circulation around all other excised circles $\partial B_k$ for $k\neq j$, 
\begin{equation}\label{eq:circcond}
\int_{\partial B_j}\frac{d v_i}{dz}dz=-\delta_{ij}.
\end{equation}
\citet[pp. 67]{crowdy2020solving} showed that in the canonical domain, $v_j(z)$ can be written in terms of the prime function as follows,
\begin{equation}\label{eq:vdef}
v_j(z)=\frac{1}{2\pi \mathrm{i}}\log{\left(\frac{\omega\left(z,\theta_j\left(1/\overline{a}\right)\right)}{\omega\left(z,1/\overline{a}\right)}\right)}-\frac{1}{2\pi \mathrm{i}}\log{\left(\frac{-\left(1-\overline{\delta_j}/\overline{a}\right)}{q_j}\right)}+v_j(1/\overline{a})+\frac{\tau_{jj}}{2},
\end{equation}
where $\theta_j$ is the M{\"o}bius map defined by,
\begin{equation}\label{eq:mobius}
\theta_j(z)=\delta_j+\frac{q_j^2 z}{1-\overline{\delta_j}z},
\end{equation}
and $\tau_{jj}$ is a constant element of the so-called period-matrix as defined by \citep[pp.31]{crowdy2020solving}. In our developments, the value of $\tau_{jj}$ is immaterial. Since our problem requires a potential with constant \emph{real} part, the functions $\mathrm{i}v_j$ are of interest since they possess constant real part on each boundary. Note that an induced charge exists on the boundary $\partial B_j$, for the functions $\mathrm{i}v_j$, rather than a circulation as in (\ref{eq:circcond}). This is consistent with the fact that the electrostatic potential, $V=\mathrm{Re}\left\{ W_e \right\}$, must be single-valued function.

Note also that $v_j$ is independent of the choice of $a$ despite the appearance of (\ref{eq:vdef}). For our purposes, we can eliminate the constants in (\ref{eq:vdef}) and define the new functions
\begin{equation}\label{eq:vtildef}
\tilde{v}_j(z)=\frac{1}{2\pi \mathrm{i}}\log{\left(\frac{\omega\left(z,\theta_j\left(1/\overline{a}\right)\right)}{\omega\left(z,1/\overline{a}\right)}\right)}-\frac{1}{2\pi \mathrm{i}}\log{\left(\frac{-\left(1-\overline{\delta_j}/\overline{a}\right)}{q_j}\right)}.
\end{equation}
Note that $\tilde{v}_j(z)$ in (\ref{eq:vtildef}) can be computed directly using the numerical implementation of the prime function developed by \cite{crowdy2007computing}.

\subsection{Solution Procedure}
The solution for the $N-$ body problem can clearly be represented as a linear superposition of the $v_j$ plus a constant. That is,
\begin{equation}\label{eq:supepos}
W_E=\alpha_0+\mathrm{i} \sum_{j=1}^{N-1}\alpha_j \tilde{v}_j(z),
\end{equation}
where $\alpha_j$ are $N$ real constants and $j\in \{0,1,...,N-1\}$. The boundary conditions on the $N$ bodies fix the $N$ undetermined coefficients.

The electrical potential is specified on each surface so that $\mathrm{Re}\left\{W_E\right\}=V_i$ for $z\in \partial B_i$, where $\partial B_N$ indicates the unit circle and $\partial B_j$ for $j\in\{1,2,...,N-1\}$ are the $N-1$ excised circles. We define the $N-$dimensional vector $\boldsymbol{V}$ whose $j^{\mathrm{th}}$ component is $V_j$ for $j=\{1,...,N\}$. We also define the $N-$dimensional coefficient vector $\boldsymbol{\alpha}$ defined by,
\begin{equation}
\boldsymbol{\alpha} = \left[
\begin{array}{cccc}
\alpha_0&\alpha_1&\cdots&\alpha_{N-1}
\end{array}  \right].
\end{equation}
Lastly, we define the matrix $\boldsymbol{A}$ with components $A_{i0}=1$ and $A_{ij}=-\mathrm{Im}\left\{\tilde{v}_j(\delta_i+q_i)\right\}$ for $j>0$. Then the coefficients $\boldsymbol{\alpha}$ are determined by the following linear algebra problem,
\begin{equation}\label{eq:linalgp}
\boldsymbol{A}\boldsymbol{\alpha}^T=\boldsymbol{V},
\end{equation}
which can be solved using the backslash operator in MATLAB.
\subsection{Floating conductors}
Suppose now that the votlage on the $p^{\mathrm{th}}$ conductor, $\partial B_p$, is unspecified. The floating conductor, if uncharged, must satisfy the condition 
\begin{equation}
\mathrm{Im}\left\{\int_{\partial B_p}\frac{dW_E}{dz} dz\right\} =0,
\end{equation}
which implies that $\alpha_p=0$. Thus, to enforce the floating boundary condition on the $p^{\mathrm{th}}$ conductor, the $p^{\mathrm{th}}$ term of the summation in (\ref{eq:supepos}) must be deleted. The corresponding linear algebra problem in (\ref{eq:linalgp}) has its $p^{\mathrm{th}}$ row removed. The value of $V_p$ is then determined through the solution of the new linear algebra problem. If $M$ conductors have floating voltages, the summation in \ref{eq:supepos} has the corresponding $M$ terms removed and the 
linear algebra problem (\ref{eq:linalgp}) has the corresponding $M$ rows removed.

\newpage

\bibliographystyle{jfm}
\bibliography{manuscript_text}

\end{document}